\documentclass[iop]{emulateapj}
\usepackage{natbib,amssymb,amsmath,graphicx,here}
\newcommand\tna{\,\tablenotemark{a}}
\newcommand\tnb{\,\tablenotemark{b}}

\begin{document}

\title{Adaptive Optics imaging of VHS~1256-1257: A Low Mass Companion to
a Brown Dwarf Binary System}

\author{Jordan M. Stone\altaffilmark{1}}
\altaffiltext{1}{Steward Observatory,
University of Arizona,
933 N. Cherry Ave,
Tucson, AZ 85721-0065 USA} 
\author{Andrew J. Skemer\altaffilmark{1,2}}
\altaffiltext{2}{Department of Astronomy and Astrophysics, 
University of California, Santa Cruz, 
1156 High St, 
Santa Cruz, CA 95064, USA}
\author{Kaitlin M. Kratter\altaffilmark{1}}
\author{Trent J. Dupuy\altaffilmark{3}}
\altaffiltext{3}{The University of Texas at Austin, 
Department of Astronomy, 
2515 Speedway C1400, 
Austin, TX 78712, USA}
\author{Laird M. Close\altaffilmark{1}}
\author{Josh A. Eisner\altaffilmark{1}}
\author{Jonathan J. Fortney\altaffilmark{2}}
\author{Philip M. Hinz\altaffilmark{1}}
\author{Jared R. Males\altaffilmark{1}}
\author{Caroline V. Morley\altaffilmark{2}}
\author{Katie M. Morzinski\altaffilmark{1}}
\author{Kimberly Ward-Duong\altaffilmark{4}}
\altaffiltext{4}{School of Earth and Space Exploration, Arizona State
University, Tempe, AZ, 85287, USA}

\begin{abstract} 

Recently, Gauza et al. (2015) reported the discovery of a companion to the late
M-dwarf, VHS~J125601.92-125723.9 (VHS~1256-1257). The companion's absolute
photometry suggests its mass and atmosphere are similar to the HR 8799 planets.
However, as a wide companion to a late-type star, it is more accessible to
spectroscopic characterization. We discovered that the primary of this system
is an equal-magnitude binary. For an age $\sim300$ Myr the A and B components
each have a mass of $64.6^{+0.8}_{-2.0}~M_{\mathrm{Jup}}$, and the b component
has a mass of $11.2^{+9.7}_{-1.8}$, making VHS~1256-1257 only the third brown
dwarf triple system. There exists some tension between the spectrophotometric
distance of $17.2\pm2.6$ pc and the parallax distance of $12.7\pm1.0$ pc.  At
12.7 pc VHS~1256-1257 A and B would be the faintest known M7.5 objects, and are
   even faint outliers among M8 types.  If the larger spectrophotmetric
distance is more accurate than the parallax, then the mass of each component
increases.  In particular, the mass of the b component increases well above the
deuterium burning limit to $\sim35~M_{\mathrm{Jup}}$ and the mass of each
binary component increases to $73^{+20}_{-17}~M_{\mathrm{Jup}}$.  At
   17.1 pc, the UVW kinematics of the system are consistent with membership in
      the AB~Dor moving group. The architecture of the system resembles
a hierarchical stellar multiple suggesting it formed via an extension of the
star-formation process to low masses.  Continued astrometric monitoring will
resolve this distance uncertainty and will provide dynamical masses for a new
benchmark system.  

\end{abstract}

\section{Introduction} 

Brown dwarfs in the field follow a tight sequence in near-infrared color
magnitude diagrams \citep[e.g.,][]{Dupuy12}. At the hottest and most massive
end, M-type dwarfs transition to L-type dwarfs as they cool. L-dwarfs are
characterized by red color and thick clouds with CO as the dominant carrier of
atmospheric carbon. At $T_{\mathrm{eff}}\sim1300~\mathrm{K}$, the L-dwarfs
undergo a dramatic transition to T-type, becoming relatively cloud free,
bluer, and methane dominated.

\begin{figure}
\plotone{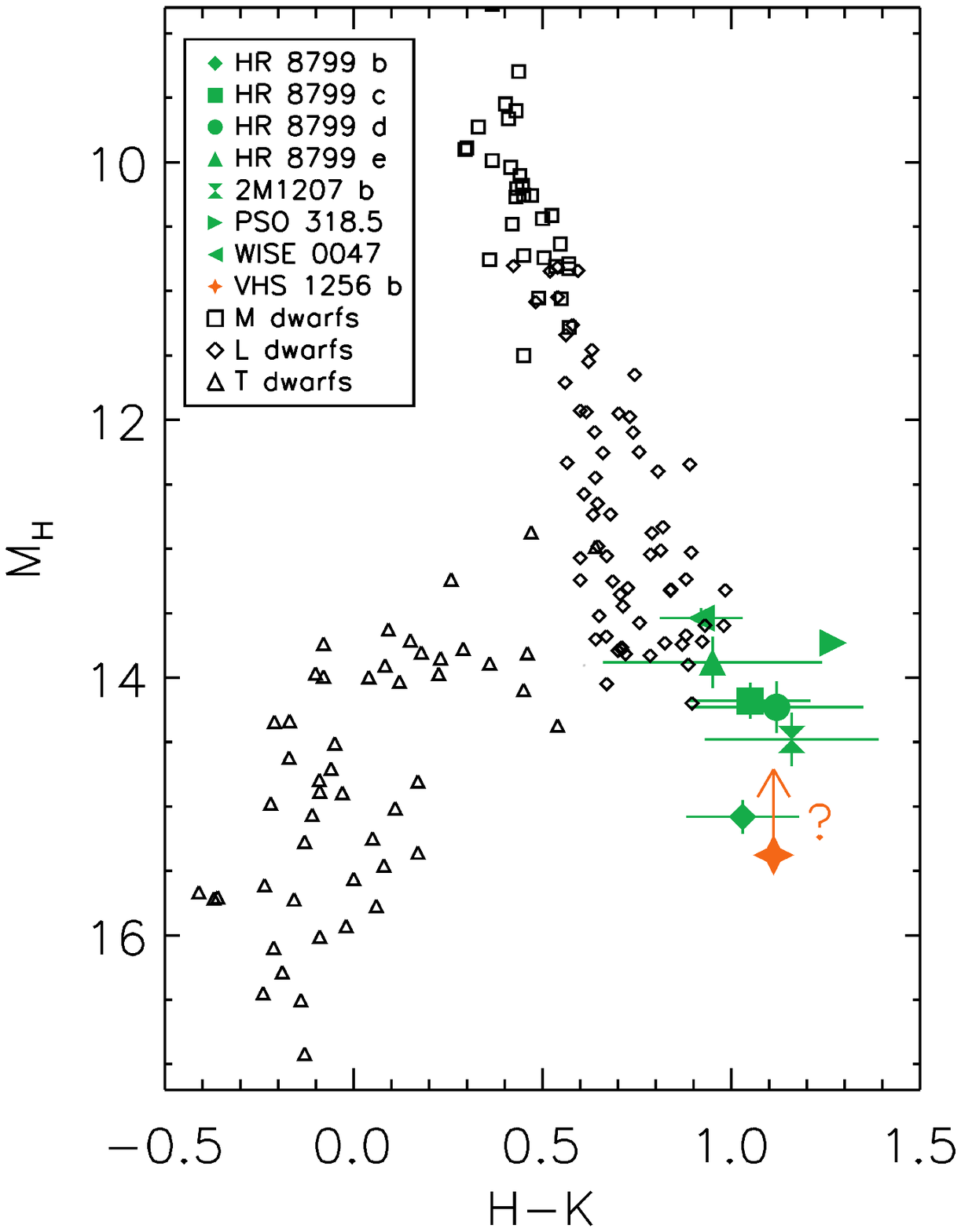}
\caption{Near-infrared color magnitude diagram. The field brown dwarf sequence
is shown with open symbols \citep[data from][]{Dupuy12}. The L to T transition
begins at M$_{\mathrm{H}}\sim14$. We also plot, with green filled symbols, the
planets in the HR~8799 system \citep[data compiled
from][]{Marois08,Metchev09,Marois10,Skemer12}, the planetary mass companion
2M~1207~b \citep[$M\sim8~M_{\mathrm{Jup}}$,][]{Chauvin04}, and the
free-floating object PSO~J318.5 \citep{Liu13}. With an orange star we show
VHS~1256-1257~b, which appears furthest from the field-dwarf sequence. The
binary nature of its host may imply a larger distance to the system which would
push VHS~1256-1257~b up in the diagram.\label{cmd}} 
\end{figure}

In the last few years, a growing population of objects at effective
temperatures where field brown dwarfs are seen to be T-type, are instead
observed to be L-type, suggesting an extension of the L-dwarf sequence to low
temperature (Figure \ref{cmd}). These objects include the four directly imaged
planets in the HR~8799 system \citep{Marois08,Marois10}, the planetary mass
binary companion 2MASS~1207~b \citep{Chauvin04}, and free floating brown dwarfs
such as the planetary mass PSO~J318.5-22 \citep{Liu13}.  The common
characteristic of these objects is low gravity due to low mass and young age.

A recently discovered companion, VHS~J125601.92-125723.9~b \citep[hereafter
VHS~1256-1257~b, $d=12.7~\mathrm{pc}$, $M\sim11M_{\mathrm{Jup}}$, age 150--300
Myr;][]{Gauza15}, is thought to be one of the most extreme members of the
extension to the L-dwarf sequence (Figure \ref{cmd}).  It appears in many ways
to be an analogue of HR~8799~b, the faintest and least massive planet in the
HR~8799 system.  This discovery is significant because VHS~1256-1257~b is
substantially easier to observe than HR~8799~b as it orbits a fainter host at
larger separation.  This means that more precise measurements can be made with
a wider array of instruments because issues with high contrast are alleviated.

The variety of objects that occupy the extended L-dwarf region in
color-magnitude diagrams suggests that distinguishing planets, which form via
accumulation of material in a circumstellar disk \citep[e.g.,][]{Sallum15},
from brown dwarfs, which form as an extension of the star-formation process to
low masses \citep[e.g.,][]{Chabrier14}, requires some care; additional
constraints such as compositional differences \citep[e.g.,][]{Skemer15}, or
system architecture \citep[e.g.,][]{Lodato05} must be incorporated into the
analysis.

In this letter we present evidence that the host of VHS~1256-1257~b is
actually a $0.1''$ equal-mass brown dwarf binary system and discuss the
implications for the nature of this system\footnote{Note added in proof:
independant ao-images presented by Rich et al., in prep}.

\section{Observations and Results } \label{ObsSec}

\subsection{MagAO/Clio2 L'} On 2015 June 4 UT, we imaged the VHS~1256-1257
system (all components simultaneously) with MagAO/Clio2
\citep{Sivanandam06,Hinz10,Close13} as part of a campaign to characterize the
atmospheres of low-gravity cool companions to nearby stars. The PSF of the
primary appeared extended compared to brighter point-sources observed the same
night. 

Our images were obtained as acquisition images for a spectroscopic observation,
so they were not optimized for highest spatial resolution. We observed in
coarse platescale mode \citep[$27.477\pm0.085$ mas
pixels$^{-1}$;][]{Morzinski15} through the L' filter. We used 0.28~s
integrations to keep the bright sky emission in the linear range of the
detector.  Files were written with five coadded frames each.  We nodded the
telescope in an AB pattern, saving ten files per position.  From each of the
ten images per nod, we subtracted the median of the ten frames from the
opposite nod position. This corrects for the bright sky emission, dark current,
and the detector bias.

\begin{figure} 
\plotone{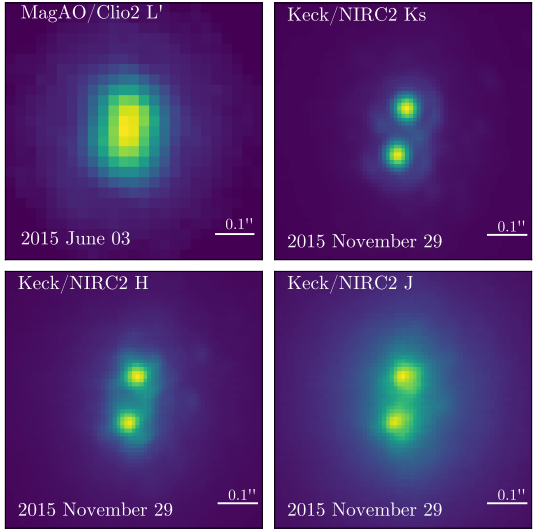} 
\caption{Images of the central binary in the VHS~1256-1257 system. North is up,
and East is Left. The cool companion is 8'' to the southwest, off the
frame.\label{binaryFig}} 
\end{figure}

We measured the flux ratio, separation, and position angle of the
marginally-resolved binary components in each of our images by fitting
a two-component model created using an image of HIP~57173 ---an unresolved
star--- as a template PSF.  Our fitting procedure optimized 6-parameters
including an amplitude and an $(x,y)$ position for each source. The best-fit
model gave a magnitude difference of $0.07\pm0.03$ a separation of $109\pm1.8$
milliarcseconds and a PA of $173\fdg3\pm0\fdg9$.  Errorbars were derived by
bootstrapping the final stack of background subtracted images.  All of our
deduced binary parameters are reported in Table \ref{table}. Our final stacked
image is shown in Figure \ref{binaryFig}.

\subsection{Keck NIRC2 J, H, Ks}
To confirm the binary nature of VHS~1256-1257~AB we conducted follow-up imaging
with a larger aperture and at shorter wavelengths to achieve better spatial
resolution. We used natural-guidestar adaptive optics \citep{Wizinowich00} and
the NIRC2 instrument on the Keck II telescope on 2015 November 29 UT. Data were
collected through the J, H, and Ks filters, and the camera was set up in the
9 mas pixel$^{-1}$ mode.  We used a three-point dither pattern in order to
track variable sky background. Files were saved using three coadds of five
second exposures.  We obtained a total of 45~s of open shutter time per filter.

We reduced and analyzed our NIRC2 images in the same fashion as in
previous work \citep[e.g., see][]{Dupuy2009, Dupuy2015c}.  For
each individual image, we applied standard calibrations (dark
subtraction, flat fielding) and then fit an analytic, three-component
Gaussian model to both objects simultaneously.  To convert the raw
$(x,y)$ coordinates from these fits to separation and position angle
(PA) on the sky, we applied the nonlinear distortion solution of
\citep{Yelda2010} and their corresponding pixel scale of
$9.952\pm0.002$\,mas\,pixel$^{-1}$ and NIRC2 header orientation
correction of $+0\fdg252\pm0\fdg009$.  In Table~\ref{table} we report the
mean and rms of the binary parameters we derived from our individual
images.  The separation measurements across $JHK_S$ bands are in good
agreement within the quoted uncertainties, but the PA values do not
agree within their much smaller quoted uncertainties.  Since the
binary components have nearly identical fluxes and colors, this is
unlikely to be due to a chromatic effect, but it could be caused by
recent changes to the nonlinear distortion of NIRC2 (J. Lu, 2015
private communication).  We therefore adopt the mean and rms of the
separations and PAs determined across $JHK_S$ bands as the final
values, $123.6\pm0.4$\,mas and $170\fdg2\pm0\fdg6$, respectively.

We also briefly observed the outer tertiary component, VHS~1256-1257~b. We
obtained three frames in three separate nod positions--- one of which contained
all three components at the same time. An exposure of 60 seconds was collected
for each nod. We reduced these images as explained above. The object appears
single down to our sensitivity. We see no equal magnitude component down to
$\sim70$~mas and no component 1 magnitude fainter down to $\sim100$~mas.  We
cannot rule out companions more than 2 magnitudes fainter than VHS~1256-1257~b.

\subsection{Ruling Out the Background Hypothesis} 

To confirm that the binary components are physically related and not the result
of a chance alignment, we inspected 2MASS images from 1999 March 1 UT. We see
no source at the present-day location of VHS~1256-1257 AB, and no sources with
the correct magnitude within 2 arcminutes. Given the high proper motion of the
system ($\sim300$ mas yr$^{-1}$) this strongly suggests the pair is a common
proper motion binary and is physically bound.

\begin{deluxetable*}{llllllll}
\tabletypesize{\scriptsize}
\tablecolumns{8}
\tablewidth{0pt}
\tablecaption{Binary Properties}
\tablehead{
    \colhead{Date}                 &
    \colhead{Filter}               &
    \colhead{Flux Ratio}           &
    \colhead{Separation}           &
    \colhead{PA}                   &
    \colhead{m$_{\mathrm{A}}$\tna}                &
    \colhead{m$_{\mathrm{B}}$\tna}                &
    \colhead{d$_{L}$\tnb}          \\ 
    \colhead{[UT]}                 &
    \colhead{}                     &
    \colhead{[mag]}                &
    \colhead{[mas]}                &
    \colhead{[$^{\circ}$]}                &
    \colhead{[mag]}                &
    \colhead{[mag]}         &
    \colhead{[pc]} }
\startdata
2015 Nov 29 & J  & $0.05\pm0.04  $ & $123.1\pm1.0$ & $170.8\pm0.2$ & $11.76\pm0.05$ & $11.78\pm0.05$ & $16.8\pm2.4$\\ 
2015 Nov 29 & H  & $0.04\pm0.02$ & $123.8\pm0.4$ & $170.1\pm0.1$ & $11.21\pm0.05$ & $11.24\pm0.05$ & $17.4\pm2.6$   \\ 
2015 Nov 29 & Ks & $0.04\pm0.03$ & $123.9\pm0.5$ & $169.6\pm0.1$ & $10.79\pm0.04$ & $10.81\pm0.04$ & $17.2\pm2.6$ \\
2015 Jun 03 & L' & $0.07\pm0.03  $ & $109.4 \pm 0.9$ & $173.3   \pm0.9$ &\nodata&\nodata&\nodata\\
\enddata
\tablenotetext{a}{Individual magnitudes determined using 2MASS unresolved
photometry and our flux ratio. We assume the J- and H-band flux ratios we
measure in the MKO system will be the same in the 2MASS system given the
similar spectral types of each component implied by the flux ratio being so
close to one.}
\tablenotetext{b}{Spectrophotometric distance implied by M7.5 spectral type.
Uncertainty includes contributions from measurement errors and the rms
variation in the population of well characterized sources in \citet{Dupuy12}.}
\label{table}
\end{deluxetable*}

\section{Discussion and Analysis} \label{DiscSec} 

\subsection{Distance and Luminosity}
\begin{figure}
\plotone{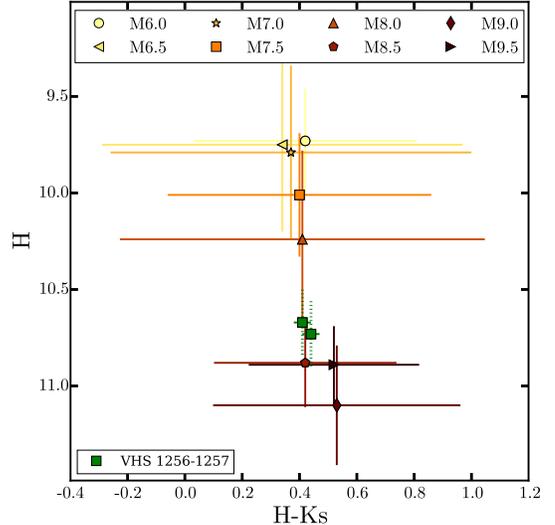} 
\caption{Near-infrared color magnitude diagram of late-M type objects in the
field. Data taken from \citet{Dupuy12} show the average and rms spread for each
spectral type. We also show VHS~1256-1257 A and B (both spectroscopically
determined to be M7.5$\pm$0.5) with the absolute magnitude implied given the
distance of 12.7 pc reported by \citet{Gauza15}.  VHS~1256-1257 A and B are
more than 2-$\sigma$ less luminous than average field M7.5 objects and are
1-$\sigma$ less luminous than typical M8s. The tension with expectations is
amplified by the age of the system because young objects like VHS~1256-1257 are
expected to be more luminous than field objects, not less. \label{primarycmd}}
\end{figure}

\begin{figure}
\plotone{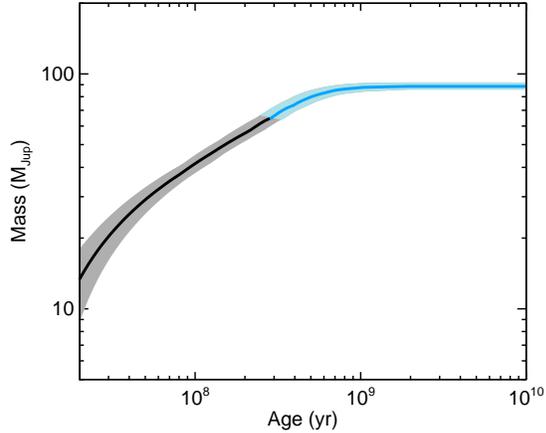} 
\caption{Age versus mass at constant luminosity according to the evolutionary
model grid of \citet{Chabrier2000}. We assume the luminosity of each binary
component if the parallax distance reported by \citet{Gauza15} is correct.
The width of the curve accommodates the uncertainty in the luminosity from our
reported photometric errors and the error in the distance measurement. We
highlight in blue models with less than 10\% of the primordial lithium
abundance, which are consistent with the non-detection from
\citet{Gauza15}.\label{newMass}}
\end{figure}

\citet{Gauza15} report a parallax distance to VHS~1256-1257 of $12.7\pm1$~pc
using $\sim9$ months of astrometric monitoring data ---collected in a variety
of filters--- together with a 2MASS point.  This distance was consistent with
the spectrophotometric distance of the central source, given the
spectroscopically determined type of M7.5$\pm$0.5 \citep{Gauza15},
assuming that it was a single star with typical absolute magnitude
\citep[averages taken for field objects from][]{Dupuy12}.  However, the fact
that the source is an equal magnitude binary, not a single star, requires
a reexamination of the spectrophotometric distance. An equal magnitude binary
will be 0.75 magnitudes brighter and the derived spectrophotometric distance
will scale by a factor of $\sqrt{2}$. In Table~\ref{table}, we show the implied
spectrophotometric distance of VHS~1256-1257 from the apparent magnitudes of
the resolved components. We use the average and rms scatter for M7.5 spectral
type objects with precise parallax measurements from \citet{Dupuy12}. The
distances derived from each band are consistent with each other and suggest
that the system is located at $\sim17.1\pm2.5$~pc, with the uncertainty
dominated by the rms variation in absolute magnitude of the population.

In this case, the distance modulus to VHS~1256-1257 increases by 0.66 mags
and components move up in color magnitude-diagrams. In particular, the
planetary mass companion would appear more like the hotter, more-massive
HR~8799~d than HR~8799~b (Figure~\ref{cmd}). Furthermore, a higher luminosity
for VHS~1256-1257~b also implies a higher mass ---up to $35~M_{\mathrm{Jup}}$
\citep[for 300 Myr, $M_{Ks}=13.38$;][]{Baraffe15}--- placing it above the
deuterium burning limit.  

The UVW kinematics of the system also change if it is located at 17.1~pc.
While \citet{Gauza15} show that the UVW velocities of VHS~1256-1257 are
consistent with the $\beta$~Pic moving group, they argue that it cannot be
a member due to the lack of lithium and the much older age.  We made use of
BANYAN II tool \citep{Malo2013, Gagne2014} to determine the probability that
the system is a member of nearby moving groups. We indicated that VHS~1256-1257
is less than 1 Gyr old based on the very-low gravity designation of the
b component.  At a distance of 17.1~pc the BANYAN II predicts membership in the
older AB Dor moving group with 66.85\% probability, membership in the young
field population with a 32.52\% probability, and membership in the $\beta$~Pic
moving group of 0.48\%.

This scenario requires the parallax for the system to be less precise than
reported, possibly due to the inhomogeneous nature of the astrometric data and
the small number of epochs used to disentangle the parallax motion from the
proper motion.  Alternatively, the parallax distance is correct and the
absolute magnitude of the binary components are intrinsically quite low. 

In Figure \ref{primarycmd}, we show the average and rms spread in absolute
H-band magnitude and H-Ks color for late-M field brown dwarfs \citep[data
from][]{Dupuy12}. VHS~1256-1257 A and B lie well below the average M7.5 point
by more than 2 $\sigma$. The uncertainty in the spectral type of the
VHS~1256-1257~A and B does accommodate a fainter M8 spectral type. Even for
this spectral type the binary components are less luminous than expected, yet
are near the edge of the rms scatter in the well characterized field
population. At Ks band, if the system is at 12.7~pc, VHS~1256-1257~A and B 
are the least luminous M7.5 objects known with M$_{\mathrm{Ks}}=10.26\pm0.18$.
GRH~2208-20, an M7.5 object, has M$_{\mathrm{Ks}}=10.11\pm0.06$, all other M7.5
sources with precise parallax have M$_{\mathrm{Ks}}<10$ \citep{Dupuy12}. Even
among M8 types, at 12.7~pc VHS~1256-1257~A and B are faint. The only other M8
with an absolute K-band luminosity greater than 10 is LHS~2397a~A
(M$_{\mathrm{Ks}}=10.13\pm0.07$).

Such low luminosities are odd given age constraints inferred from the INT-G and
VL-G gravity designations given to the unresolved binary and the companion
\citep[age $<500$ Myr;][]{Gauza15}.  Young low-gravity objects are expected to
be more luminous than field objects based on theoretical evolution models and
observations of young populations \citep[e.g.,][]{Casewell07,Baraffe15}.
However, the number of well characterized INT-G M7-8 objects with which to
compare our measured absolute magnitudes is small, making it difficult to rule
out the low-luminosity explanation.

\subsection{Mass and Orbital Period}

If the parallax distance is correct, then the mass of the binary components
need to be updated. Using the evolutionary models of \citet{Chabrier2000} we
construct a plot of age versus mass at constant luminosity (Figure
\ref{newMass}). We assume a luminosity for each source of
$log(\frac{L}{L_{\odot}})=-3.44\pm0.1$, half that reported by \citet{Gauza15}.
Since the objects have no detectable lithium \citep{Gauza15}, we indicate in
Figure \ref{newMass} those models which have less than 10\% of the primordial
lithium abundance by coloring them blue.  The minimum mass for each of the
binary components is $64.6^{+0.8}_{-2.0}~M_{\mathrm{Jup}}$ and the
corresponding minimum age is $280^{+40}_{-50}$~Myr. 

Using our new mass estimate, we can calculate an orbital period for
VHS~1256-1257~A and B.  To do this, we take the projected separation measured
in our discovery image and correct it using the projected
separation--semi-major axis correction factor for a moderate discovery bias
(due to the close separation of the binary) taken from \citet{Dupuy11}. We find
an orbital period of $5.87\pm2.7$~yr. If the system is located at 17.1~pc, as
suggested by the photometry of the components, then the period changes because
the projected separation increases as does the mass of each component. In this
case, the orbital period is $8.7\pm4.3$ yr. Given these orbital periods and our
observation of $\lesssim3^{\circ}$ of change in the PA of the binary
components, the binary orbit cannot be face-on and circular.

\subsection{Formation and Dynamical Evolution} 

We have shown that VHS~1256-1257 consists of three brown dwarf objects
organized into a hierarchical triple system, the A and B components being
orbited by the more distant b component.  Only two other brown
dwarf triples appear in the literature, 2MASS J08381155+1511155 \citep[composed
of three T-dwarfs;][]{Radigan2013}, and DENIS-P J020529.0-115925
\citep[marginally resolved late-L and early T components;][]{Bouy2005}. Neither
includes M-type objects nor a planetary mass component. 2M0441+2301 AabBab
\citep{Todorov2010, Bowler2015} includes three substellar components including
one below the deuterium burning limit, yet the system also includes a stellar
component and the hierarchical quadruple architecture is much different from
VHS~1256-1257. VHS~1256-1257 shares many characteristics with 2MASS
J01033563-5515561(AB)b, a hierarchical triple with a planetary mass companion
orbiting a binary composed of M6-type stars \citep{Delorme13}. However,
VHS~1256-1257 is closer, older, and less massive.

The hierarchical-triple orbital configuration of VHS~1256-1257 suggests that
the system is more akin to a low mass analog of a stellar multiple, rather than
a planetary system with a planet on a wide orbit.  Models for turbulent
fragmentation suggest that isolated objects may form at masses more than
a factor of two below VHS 1256-1257~b \citep{Hennebelle:2008}.  The separation
of VHS 1256-1257 AB is consistent with the sample of brown dwarf binaries
observed by \citet{Close03} and \citet{Dupuy:2011}, who showed that low mass
binaries typically have smaller semi-major axes than their stellar
counterparts. These scaled down separations are expected if brown dwarfs form
as a low mass tail of the turbulent core fragmentation process
\cite{Jumper:2013}. The presence of a high mass ratio ($\sim$10\%) tertiary at
relatively wide separations is similarly consistent with low mass star
formation \citep{Reipurth:2001, OKMKK10, Bate:2012}.  If the three objects
formed from the same filament, the objects could have undergone early dynamical
interactions which naturally tighten one orbit in exchange for softening the
outer most orbit \citep{Heggie:1975, Reipurth15}.  We suggest early dynamical
interactions are the most likely origin for the orbital configuration due to
low stellar densities in the solar neighborhood.  Although mutual inclination
between the two orbits could induce Kozai-Lidov oscillation on Myr timescales,
the inner orbit is too wide to undergo significant tidal evolution over the
lifetime of the system \citep{Fabrycky:2007}.

\section{Conclusion}

We have revealed that the host of VHS~1256-1257~b is actually an equal-mass
brown dwarf binary, making VHS~1156-1257 the third triple system known with
exclusively substellar components. There is some tension between the parallax
distance and the spectrophotometric distance to the system.  This implies that
either the system is more distant than reported, and the planetary mass
companion more massive ($\sim35~M_{\mathrm{Jup}}$ for an age of 300~Myr) or the
components are less luminous than expected based on their age and spectral
type. The architecture of this system is consistent with outcomes from normal
low-mass star formation. Continued astrometric monitoring of this system will
resolve the tension between the luminosity and parallax distances, and enable
dynamical mass measurements for the components, providing important benchmarks
for these low-gravity objects.

\acknowledgements{Some of the data presented herein were obtained at the W.M.
Keck Observatory, which is operated as a scientific partnership among the
California Institute of Technology, the University of California and the
National Aeronautics and Space Administration. The Observatory was made
possible by the generous financial support of the W.M. Keck Foundation. The
authors wish to recognize and acknowledge the very significant cultural role
and reverence that the summit of Mauna Kea has always had within the indigenous
Hawaiian community.  We are most fortunate to have the opportunity to conduct
observations from this mountain.  This work was supported by NASA Origins
grants NNX11AK57G, NNX13AJ17G and NSF AAG grant 121329. J.M.S. was also
partially supported by the state of Arizona Technology Research Initiative Fund
Imaging Fellowship. A.S.  is supported by the National Aeronautics and Space
Administration through Hubble Fellowship grant HSTHF2-51349 awarded by the
Space Telescope Science Institute, which is operated by the Association of
Universities for Research in Astronomy, Inc., for NASA, under contract NAS
5-26555.} 

\end{document}